\documentclass[conference]{IEEEtran}
\IEEEoverridecommandlockouts
\usepackage{cite}
\usepackage{amsmath,amssymb,amsfonts}
\usepackage{algorithmic}
\usepackage{graphicx}
\usepackage{textcomp}
\usepackage{xcolor}
\usepackage{makecell}
\usepackage{multirow}
\usepackage[official]{eurosym}
\def\BibTeX{{\rm B\kern-.05em{\sc i\kern-.025em b}\kern-.08em
    T\kern-.1667em\lower.7ex\hbox{E}\kern-.125emX}}
\begin{document}

\title{\textit{Exploratory Study on User's Dynamic Visual Acuity and Quality Perception of Impaired Images}\\
}

\author{\IEEEauthorblockN{Jolien De Letter, Anissa All, Lieven De Marez}
\IEEEauthorblockA{\textit{Media, Innovation and Communication Technologies} \\
\textit{Ghent University}\\
Ghent, Belgium}
\and
\IEEEauthorblockN{Vasileios Avramelos, Peter Lambert, Glenn Van Wallendael}
\IEEEauthorblockA{\textit{Internet and Data Science Lab (IDLab)} \\
\textit{Ghent University}\\
Ghent, Belgium \\
glenn.vanwallendael@ugent.be}
}

\maketitle

\begin{abstract}
In this paper we assess the impact of head movement on user's visual acuity and their quality perception of impaired images. There are physical limitations on the amount of visual information a person can perceive and physical limitations regarding the speed at which our body, and as a consequence our head, can explore a scene. In these limitations lie fundamental solutions for the communication of multimedia systems. As such, subjects were asked to evaluate the perceptual quality of static images presented on a TV screen while their head was in a dynamic (moving) state. The idea is potentially applicable to virtual reality applications and therefore, we also measured the image quality perception of each subject on a head mounted display. Experiments show the significant decrease in visual acuity and quality perception when the user's head is not static, and give an indication on how much the quality can be reduced without the user noticing any impairments.
\end{abstract}


\section{Introduction}\label{Intro}
Improving the resolution of head mounted devices is currently one of the main areas of focus in the field of virtual reality (VR). This was not surprising, especially since the amount of visual information displayed by VR displays is much smaller than the human viewing capacity. VR displays deliver around $15$ pixels per degree, with a $90^{\circ}$ field of view and a fixed depth of focus of two meters \cite{Langley,Abrash}. This is in large contrast to TV screens where the resolution starts to match human perception \cite{Wallendael}. Humans are, capable of perceiving around $120$ pixels per degree, a field of view between $220^{\circ}$ and $230^{\circ}$, and a depth of focus which can vary. Although it will take some time to fully exploit the human visual system, VR experiences will certainly head towards improving in that area.

Such a poor visual system performance not only yields lower quality immersive experiences, but it is also linked to greater VR sickness, which poses a comfort problem to users \cite{Davis}. VR sickness or simulator sickness, also known as ‘Virtual Reality Induced Symptoms and Effects’ (VRISE) or cybersickness, refers to a constellation of oculomotor and nausea related symptoms that users experience during and after participating in a simulated environment  \cite{Sue,Kennedy}. Similar to motion sickness, simulator sickness is caused by a conflict between the perceived visual information and the bodily senses. Amongst the several factors which have been found to contribute to VRISE, there is also the issue of lag. Lag occurs when there is a delay or latency between the actions of a user, for instance head motion, and the system’s response~\cite{Davis}. 

Considering a bad visual performance induces simulator sickness, pursuing a better view in VR will become crucial in the future \cite{Davis}. Such a view expansion with a higher resolution and a wider field of view would, consequently, demand a substantial amount of bandwidth and storage. For this reason, blurring the points outside of the human vision range as a way to minimize streaming data will become a more evident solution. This blurring in VR could occur through foveated rendering if the VR installation is equipped with eye-tracking technology or through system-feedback on the user's position and movement \cite{Patney}. 

In the current study, we have explored the possibilities of the  latter in order to minimize streaming data. As users' bodies are frequently moving around in VR, we want to learn how substantially their head movement affects their visual acuity in such a way that they see little or no detail. The goal of the current study was to learn users' changes in (a) visual acuity performance and (b) perception of impaired image quality when their heads are stationary (static) versus when their heads are in motion (dynamic). According to our knowledge, there is no other study on dynamic visual acuity targeting the improvement of multimedia applications.

From the literature we learn that visual acuity declines when horizontal movement of the head increases due to an imperfect pursuit of eye movements, resulting in a continued image motion on the retina \cite{Horng,Demer,Miller}. For television (TV) and VR, we expected to find a similar pattern. Consequently, as users in motion will see less detail, we anticipated that they would rate the impairment of images as less perceptible than when observing statically. The main contribution of this work is the realization of the perception loss of the subjects quantitatively, when their head is in a dynamic state. This is already crucial information for being able to control the delivery of high quality video at lower bitrates in the case of a dynamic viewing environment (e.g., in VR). Additionally, the results give us an indication for potentially studying cybersickness in VR, as well as useful insights for providing quality VR experiences at lower data rates.

\section{Dynamic Visual Acuity}\label{DVA}

\subsection{Introduction}\label{DVA_Intro}

\begin{figure}[t]
	\centerline{\includegraphics[width=5cm]{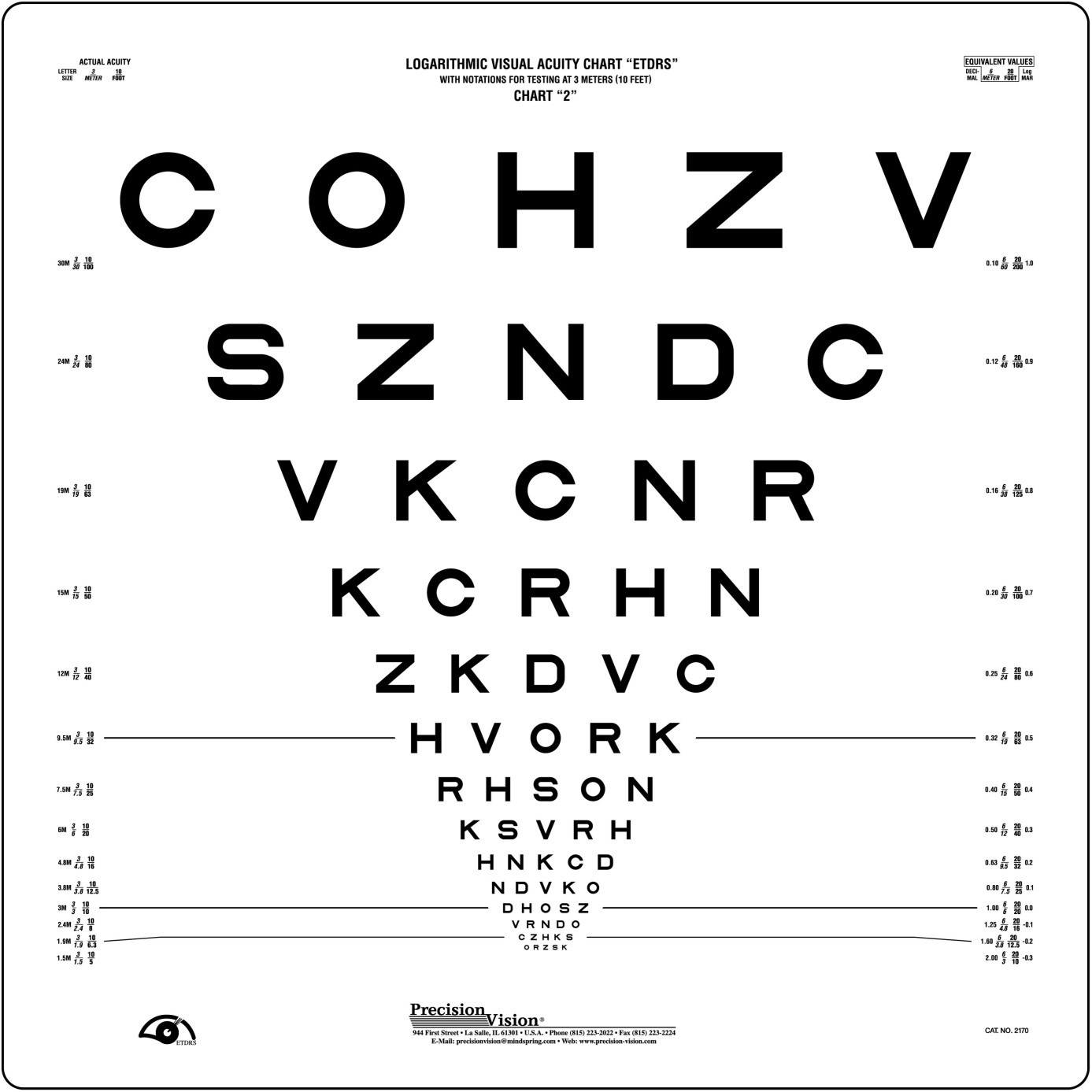}}
	\caption{Typical \textit{Sloan} visual acuity test \cite{Sloan} used for clustering the subjects and setting weights on opinion scores. It ranges between $0$ and $2$, where $2$ is the best score achieved for reading the bottom set of characters.}
	\label{fig}
\end{figure}

Visual acuity is the ability of the eye to perceive details. It depends on optical factors and neural factors. Visual acuity can be described as static and dynamic. Static visual acuity, the common measure of visual acuity, is defined as the slightest detail that the human eye can distinguish in a stationary, high contrast target (typically an eye chart with black letters on a white background). The standard \textit{Sloan} chart~\cite{Sloan} for measuring visual acuity is shown in Fig.~\ref{fig}.

Dynamic visual acuity is the ability of the eye to visually discern fine detail either in a moving object, or a stationary object while the actual subject is in motion, or both. In other words, dynamic visual acuity is the ability to distinguish fine detail when there is relative motion between the object and the observer. While static visual acuity is the common measure, for scenarios where motion is present such as driving licensing, dynamic visual acuity should also be considered. Similarly, dynamic visual acuity would seem to be more relevant when subjectively evaluating VR content, or any other content present in a dynamic viewing environment.

So far, practical dynamic visual acuity test methods are very limited and serving only medical purposes \cite{Peters, Li}. However, in the next sub-section we elaborate on the current status of the relevant research and literature (according to our findings). First, we focus on the research on visual acuity in respect of a vision research point of view, and then, we focus more on the research as perceived from a Quality of Experience (QoE) point of view.

\subsection{Related work}\label{Related}

There have been attempts in the past to increase our knowledge on perception aspects under movement both from a vision research point of view and from a QoE point of view. With respect to vision research, research has been expanded towards movement starting at the basis, our eyes. The eye is limited in its visual capacity and these limits become more stringent with increased freedom of movement. When considering the eye by itself, only the foveola (approx. $1^{\circ}$ of vision) and the fovea (approx. $5^{\circ}$ of vision) provide us with a high acuity of vision. When moving further from our gaze direction, our visual acuity almost linearly decreases up to the mid-peripheral vision (approx. $30^{\circ}$ of vision), from where visual acuity drops even faster~\cite{Anstis}. In addition to that, lateral masking phenomena make our peripheral visual acuity decline when more information is present \cite{Monti}. When the eye starts moving and exploring the scene, one talks about saccades. Express saccades take around $100$ ms and regular saccades around 150 ms corresponding to almost 14 displayed pictures considering a screen refresh rate of $90$ Hz \cite{Fischer}. Because the eye is in motion, perception loss is inevitable \cite{Schutz}. When moving our head, visual perception starts to interact with vestibular compensation. The vestibular system in the inner ear senses body motion (mainly the head), and uses this to control movement of the eye. The vestibulo-ocular reflex (VOR) moves the eyes contrary to the head, enabling gaze stabilization for both linear and angular head motion. To measure visual acuity loss between stationary head and a rotating head, different dynamic visual acuity tests have been created \cite{Brown, Demer2}. However, in this area (vision research), the state of the art has a strict focus on visual acuity rather than error visibility of visual coding related quality loss. In this research domain, most tests involve unnatural structures such as letters and symbols to measure visual acuity and discard any realistically looking scenes. Such assumptions limit the possibility to translate such research findings explicitly to the domain of visual representation and interaction.

With respect to QoE for visual representation and coding, multiple studies have tried to push the state of the art in perception in the direction of plenoptic VR content. These studies can be classified in the QoE tests purely based on graphics and 3D meshes \cite{Doumanoglou}, and the ones based on camera captured-content \cite{Recio}. For most works, the camera-captured VR content is transformed back to regular 2D video, which eliminates the interaction possibilities and the physical movements one could make \cite{Ling}. First efforts to investigate VR quality of experience are made in ITU recommendations under development such as the ITU-T \textit{SG12, Q13, G.QoE-VR} and \textit{P.360-VR}, but these are mainly limited to $360^{\circ}$ video. Irrespective of the fact that these works are limited to only a subset of the freedom to move (only rotation of the head), they are mainly focused on overall quality of experience rather than perception capabilities on the fundamental level.

Additional to the perception restrictions of moving eyes there are the physical limitations involving movement of our body. Kinematics of the human body has been extensively investigated and modelled, but never used for the investigation of dynamic visual acuity and the actual latency to achieve a gaze direction, for all possible directions. The efforts needed for predicting that latency, which is associated with a different gaze direction, can be complemented by the research on visual saliency. Saliency research tries to model the region of interest for the observer and tries to answer the question on where the observer will be looking in certain content. The state of the art in this area seems promising since it is already expanding from 2D video towards $360^{\circ}$ VR experiences\cite{Gutierrez, Sitzmann}.

\section{Methodology}\label{Method}

\subsection{Experimental setup}\label{setup}

\begin{figure}[t]
	\centerline{\includegraphics[width=3cm]{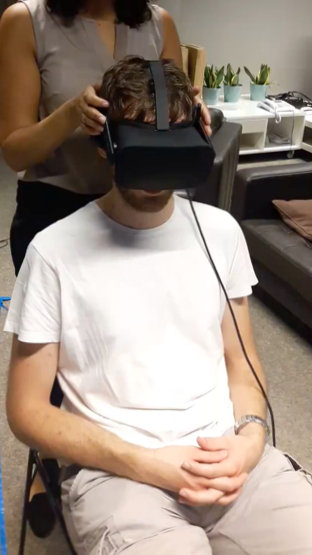}}
	\caption{The dynamic aspect of the experiment was introduced by a supervisor standing behind the subject. The participant's head was turned horizontally at a pre-determined pace of $3.5$ Hz and a width of $30^{\circ}$ \cite{Rine}.}
	\label{fig_motion}
\end{figure}

Dynamic visual acuity is the acuity which is obtained during relative motion of either the optotypes (standardized symbols for testing vision) or the observer \cite{Demer}. Contrary to the state of the art, the optotypes in this experiment remained stationary in both the dynamic and static tests. The dynamic aspect in our experiment was introduced by the subject's head movement. We chose to do this as we are interested in the visual performance of users when they move their heads, and not when the images are moving around them, because this is how viewers in VR typically interact with content. To ensure reliability, one test supervisor stood behind the subjects and turned the participants’ heads at a pre-determined pace ($3.5$~Hz) and angle ($30^{\circ}$), similar to a method described in \cite{Rine}. To facilitate this, participants remained seated during the entire duration of the experiment (see Fig. \ref{fig_motion}). One subject who felt nauseous due to VR sickness was unable to perform the dynamic tests and was, thus, excluded from further testing. In total 22 people participated. Each visual acuity test took approximately 3 to 4 minutes, while the entire test session was kept under the maximum testing limit of 30 minutes, as recommended by \cite{Keimel}.

This study used the consumer version of \textit{Oculus Rift} (https://www.oculus.com/rift) as head-mounted display to project the test interface using the \textit{Virtual Desktop} application (https://www.vrdesktop.net). The participants did not use the Touch controllers during the virtual reality projection. \textit{Oculus Rift} delivers a resolution of $1080\times1200$ pixels per eye, a frame rate of $90$ fps, and a latency between $15.3$ ms and $19.7$ ms. The TV monitor, which was used for comparison, was a \textit{Philips 9000 series Smart-LED TV} (46PFL9705k/02) with a resolution of $1280\times1024$ pixels. It had a $46$ inch diagonal dimension, and was thus satisfying the minimum requirement of $14$ inch to be used in visual quality assessment tests \cite{ITU-T}.

\subsection{Subjective test}\label{subjects}

The recruitment and the study occurred at a knowledge and innovation center in Ghent-Belgium. All participants were recruited through voluntary participation. As compensation for their time and effort, participants were given one consumption of \EUR{5} on site. In total $22$ people participated in the experiments, of which $15$ were female and $7$ were male. This exceeds the minimum-norm of $15$ as recommended by ITU-T \cite{ITU-T}. We solely recruited non-experts who were not involved in the study process, and thus, had no preconceptions about the goal of the experiment. The ages of the participants varied between $20$ and $54$ with an average age of $28.45$. If the respondents normally wore sight corrections, these were also worn during the test.

Before the actual experiments took place, a pilot study was organized. The first pilot consisted of a small scale trial run of the actual experiment in order to assess the efficacy of the experiment design, the clarity of instructions, the instruments, and the total testing time. At the start of the experiment, the subjects were informed about the procedure and they filled in a consent form. All subjects participated in a visual acuity test and an impaired image quality rating (IIQR) under four conditions: on a TV and on a head mounted display (HMD), while the user’s head was either moving or static. All subjects viewed the same set of images with counterbalanced quality levels, environment (HMD or TV), and state of the user's heads (static or dynamic). Before starting the VR test, each participant was given enough time to ensure the headset was tightly fitted around their head, and to adjust the lens slider on the bottom of the headset.

Participants were seated three meters from the TV screen. To obtain a similar distance-to-chart on the HMD screen, the display settings within \textit{Virtual Desktop} were changed to \textit{fish-eye} $30^o$. Subjects were given instructions to read each letter from the displayed \textit{Sloan} chart (see Fig.~\ref{fig}). When participants read a line incorrectly, their visual acuity value was the score of the previous line. In case participants read the last line correctly, they received the highest score possible.The \textit{Sloan} charts used for this experiment were the 2000 Series Revised ETDRS Chart 1, 2 and 3 published by Precision Vision \cite{Sloan}.

\begin{figure}[t]
	\centerline{\includegraphics[width=8.5cm]{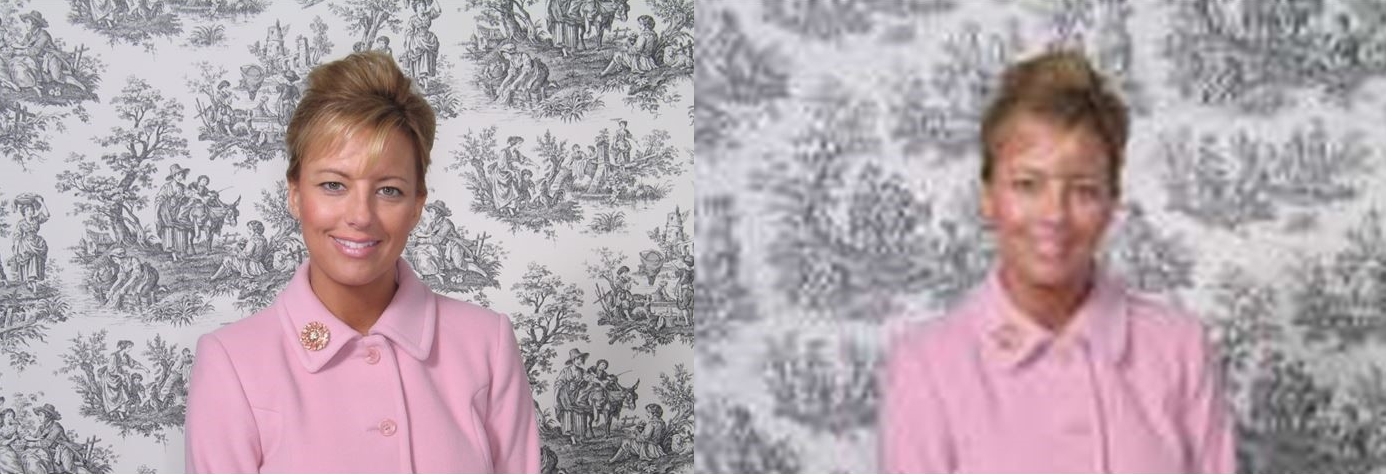}}
	\vspace{0.5cm}
	\centerline{\includegraphics[width=8.5cm]{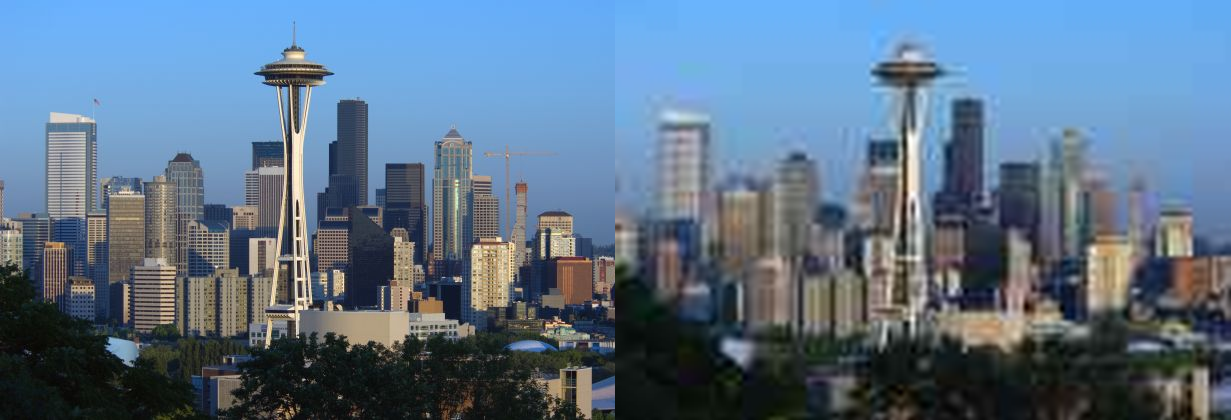}}
	\caption{Example of the evaluation method (double stimulus impairment scale) used in the actual tests. The viewer sees the original picture on the left side and the impaired picture on the right side (Top: test set \textit{woman}, bottom: test set \textit{city}).}
	\label{fig1}
\end{figure}

Each conducted impaired image quality test consisted of two phases: the training test, and the main test. The training phase introduced the subjects to the test setup and allowed them to practice the assessment tasks. The content used for the training was similar to the main test content. To keep the total testing time under $30$ minutes, we only selected two images of the same resolution ($2200\times1500$ pixels). The first image consisted of a woman's head and upper body sitting in front of a textured background, while the second image showed the landscape of a big city consisting of buildings and a blue sky as a background. The introduced impairment was a result of 1) downscaling to different resolutions and 2) JPEG coding using different quantization parameters. The evaluation technique was a Double Stimulus Impairment Scale (DSIS) method, where the reference images and the impaired images were shown side by side (see Fig.~\ref{fig1}). Subjects were aware of the position of the reference image, and they were asked to answer the following question: \textit{"How do you rate the impairment of the image on the right in comparison to the image on the left?"} on a six-point impairment scale \cite{ITU}:

\begin{enumerate}
    \item[1 -] imperceptible
    \item[2 -] just perceptible
    \item[3 -] definitely perceptible but not annoying
    \item[4 -] somewhat objectionable
    \item[5 -] definitely objectionable
    \item[6 -] very objectionable
\end{enumerate}

\noindent This is similar to the latest ITU picture assessment standardized 5-grade impairment scale \cite{ITU-R}. Each subject viewed a total of ten images per test condition: five reference images, and five impaired images in various resolutions and compression sizes of either test set $1$ (\textit{woman}) or test set $2$ (\textit{city}). Every participant received $10$ seconds to view each pair of images and assess the quality by voting from 1 to 6 according to the above described impairment scale. The average score for each case is the mean opinion score (MOS) \cite{MOS} used for visualizing the results of this work. 

\begin{table}[t]
	\centering
	\caption{Visual acuity results of the test subjects --
	Maximum possible acuity score: 2 --
	Format:~Mean(Standard Deviation).}
	\begin{center}
	\footnotesize
		\begin{tabular}{|c|c|}
		    \hline
			\textbf{Static on TV} & 1.42 (0.36) \\ \hline
			\textbf{Dynamic on TV} & 0.94 (0.36) \\ \hline
			\textbf{Static  HMD} & 1.11 (0.30) \\ \hline
			\textbf{Dynamic on HMD} & 0.93 (0.28) \\
			\hline
		\end{tabular}
		\label{tab1}
	\end{center}
\end{table}

\section{Results}

The visual acuity test scores are presented in Table~\ref{tab1}. As expected, subjects scored significantly higher on the static acuity test than on the dynamic visual acuity test both on TV and in VR. Table~\ref{tab2} summarizes our results. For each level of impairment we introduced, for every scenario (Static/Dynamic, TV/HMD), for each test set, we calculated the mean (MOS) and its standard deviation. For evaluating relevant correlations, we calculated the effect sizes between static and dynamic scenarios for each test set. Pearson's $r$ is often used as effect size when paired quantitative data are available and it is calculated as follows. Given pairs of data $(x_1,y_1),...,(x_n,y_n)$ consisting of $n$ number of pairs, the Pearson's $r$ coefficient is defined as:

\begin{equation*}
    r = \frac{{}\sum_{i=1}^{n} (x_i - \overline{x})(y_i - \overline{y})}{\sqrt{\sum_{i=1}^{n} (x_i - \overline{x})^2(y_i - \overline{y})^2}},
\end{equation*}

\noindent where $n$ is the number of the paired samples, $x_i,y_i$ are the sample points, and $\overline{x}=\frac{1}{n}\sum_{i=1}^{n}x_i$ is the mean of the samples (same for $\overline{y}$). A larger absolute value indicates a stronger effect (larger correlation).

Dynamic visual acuity scores correlated well with the dynamic IIQR results. More specifically, on TV: $r = 0.46, p < 0.05$, and on HMD: $r = 0.56, p < 0.01$), where $r$ is the correlation coefficient and $p$ is the significance level. Typically, if $p$ is lower than the conventional $5\%$ ($p < 0.05$) the correlation coefficient is called statistically significant. In other words, during the dynamic IIQR respondents with a better dynamic sight rated the impaired images as more annoying than other participants who had lower dynamic visual acuity test scores. For the static and dynamic conditions on HMD, we see a medium effect size ($r = 0.33$) at the lowest impairment level (level $1$) for the test set \textit{woman}, whereas for the test set \textit{city} there is a medium effect size ($r = 0.27$) at the impairment level $3$. When comparing to the static and dynamic conditions on TV, we validate that impairment levels $2$ and $4$ provide medium and large effect sizes for the \textit{woman} test set. Similarly, for the \textit{city} test set, the range of impairment levels $2-5$ yields medium and large effect sizes. 

\begin{figure}[t]
	\hspace{-0.2cm}
	{\includegraphics[width=8.9cm]{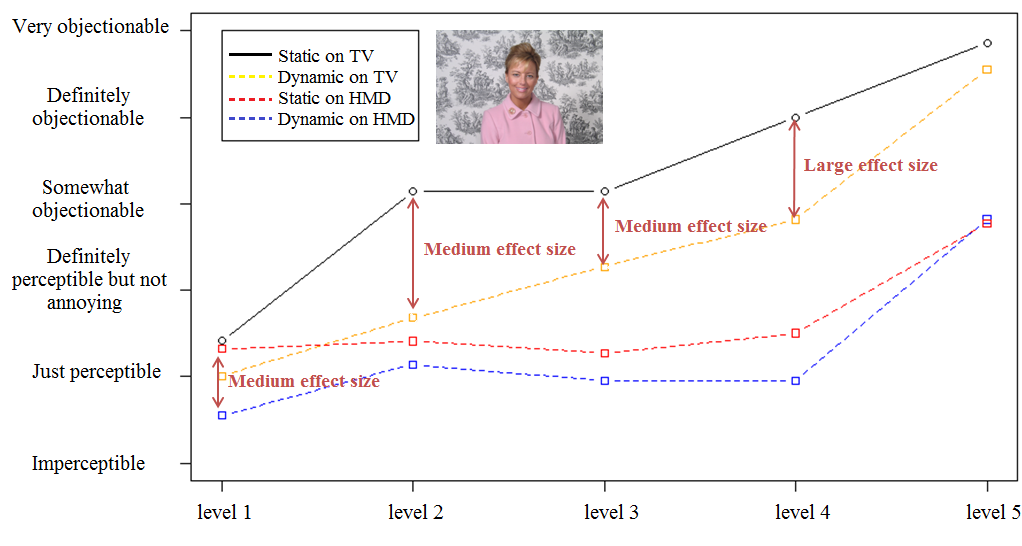}}
	\caption{Perceived impairment scale vs. level of impairment (compression/downscaling) for the test image \textit{woman}. Level 1 - 5 denotes different impairment levels scale, where level 1 is high resolution and light compression while level 5 is low resolution and heavy compression.}
	\label{fig2}
\end{figure}

\begin{figure}[t]
	\hspace{-0.2cm}
	{\includegraphics[width=8.9cm]{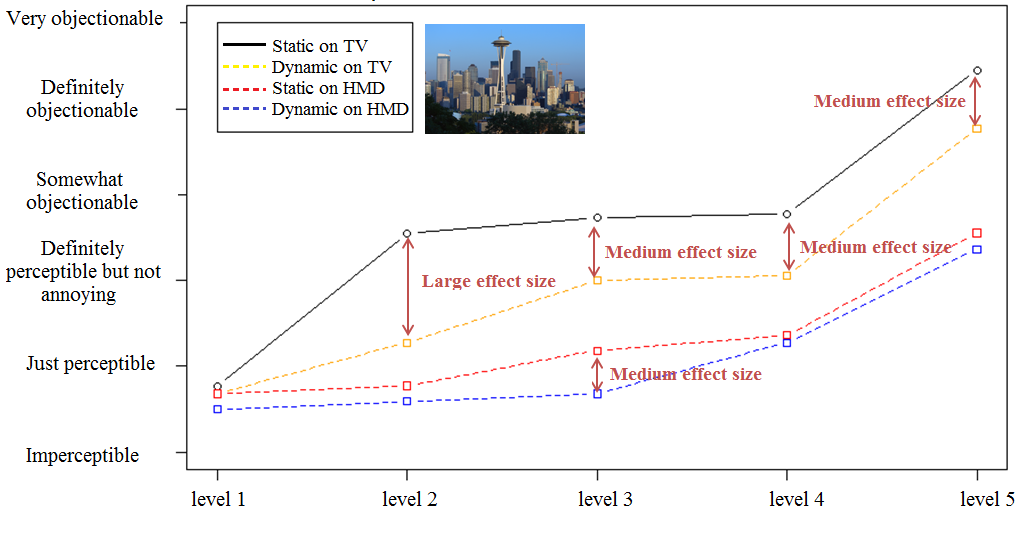}}
	\caption{Perceived impairment scale vs. level of impairment (compression/downscaling) for the test image \textit{city}. Level 1 - 5 denotes different impairment levels scale, where level 1 is high resolution and light compression while level 5 is low resolution and heavy compression.}
	\label{fig3}
\end{figure}

In Fig.~\ref{fig2} and Fig.~\ref{fig3}, those effect sizes for different impairment levels are presented in terms of the corresponding impairment scale. More explicitly, for the TV test scenario, stronger compression results in subjects perceiving more impairments. However, when the subjects are in a dynamic state, their vision is less sensitive and the impairments are less perceptible. In that way, we can get an indication of how much the quality could be reduced, without being noticeable by the majority of the subjects. For the HMD test scenario, it can be seen that only level 4 compression and higher results in impairment that gets definitely noticeable and objectionable. This might be caused by the inadequate resolution of the HMD which does not allow the user to perceive the scene in high detail. On the other hand, we can draw similar conclusion for how much the quality can be dropped (for the specific setup) without noticeable compression artifacts.



\begin{table}[]
\caption{Summarized results of the subjective test including means, standard deviations, and effect sizes (Pearson's correlation) of the mean opinion scores for different levels of impairment and different conditions.}
\resizebox{\columnwidth}{!}{%
\begin{tabular}{|c|c|c|c|c|c|}

\hline
\textbf{\begin{tabular}[c]{@{}c@{}}Impairment\\ level\end{tabular}} & \textbf{\begin{tabular}[c]{@{}c@{}}Image\\ content and \\file size\end{tabular}} & \textbf{Condition} & \textbf{Mean} & \textbf{\begin{tabular}[c]{@{}c@{}}Standard \\ deviation\end{tabular}} & \textbf{\begin{tabular}[c]{@{}c@{}}Effect size\end{tabular}} \\ \hline
\multirow{8}{*}{\begin{tabular}[c]{@{}c@{}}Level 1\\ Resolution - 616p\end{tabular}} & \multirow{4}{*}{\begin{tabular}[c]{@{}c@{}}Woman\\ 30.7 KB\end{tabular}} & Static TV & 2.53 & 0.96 & \multirow{2}{*}{0.18 (a)} \\ \cline{3-5}
& & \multicolumn{1}{l|}{Dynamic TV} & 2.12 & 1.23 & \\ \cline{3-6} 
& & Static HMD & 2.29 & 1.29 & \multirow{2}{*}{0.33 (b)} \\ \cline{3-5}
& & Dynamic HMD & 1.47 & 0.86 & \\ \cline{2-6} 
& \multirow{4}{*}{\begin{tabular}[c]{@{}c@{}}City\\ 21.7 KB\end{tabular}}  & Static TV & 1.94 & 0.61 & \multirow{2}{*}{0.06 (a)} \\ \cline{3-5}
& & Dynamic TV & 1.76 & 0.78 & \\ \cline{3-6} 
& & Static HMD & 1.82 & 0.99 & \multirow{2}{*}{0.10 (a)} \\ \cline{3-5}
& & Dynamic HMD & 1.59 & 0.74 & \\ \hline
\multirow{8}{*}{\begin{tabular}[c]{@{}c@{}}Level 2\\ Resolution - 308p\end{tabular}} & \multirow{4}{*}{\begin{tabular}[c]{@{}c@{}}Woman\\ 18 KB\end{tabular}} & Static TV & 4.00 & 1.10 & \multirow{2}{*}{0.37 (b)} \\ \cline{3-5}
& & Dynamic TV & 2.59 & 1.17 & \\ \cline{3-6} 
& & Static HMD & 2.41 & 1.18 & \multirow{2}{*}{0.11 (a)} \\ \cline{3-5}
& & Dynamic HMD & 2.18 & 1.25 & \\ \cline{2-6} 
& \multirow{4}{*}{\begin{tabular}[c]{@{}c@{}}City\\ 13 KB\end{tabular}}  & Static TV & 3.41 & 1.10 & \multirow{2}{*}{0.51 (c)} \\ \cline{3-5}
& & Dynamic TV & 2.06 & 1.03 & \\ \cline{3-6} 
& & Static HMD & 1.88 & 0.87 & \multirow{2}{*}{0.10 (a)} \\ \cline{3-5}
& & Dynamic HMD & 1.71 & 0.91 & \\ \hline
\multirow{8}{*}{\begin{tabular}[c]{@{}c@{}}Level 3\\ Resolution - 308p\end{tabular}} & \multirow{4}{*}{\begin{tabular}[c]{@{}c@{}}Woman\\ 11.7 KB\end{tabular}} & Static TV & 4.00 & 1.17 & \multirow{2}{*}{0.36 (b)} \\ \cline{3-5}
& & Dynamic TV & 3.06 & 1.08 & \\ \cline{3-6} 
& & Static HMD & 2.41 & 1.16 & \multirow{2}{*}{0.14 (a)} \\ \cline{3-5}
& & Dynamic HMD & 1.76 & 1.05 & \\ \cline{2-6} 
& \multirow{4}{*}{\begin{tabular}[c]{@{}c@{}}City\\ 8.85 KB\end{tabular}}  & Static TV & 3.59 & 1.03 & \multirow{2}{*}{0.33 (b)} \\ \cline{3-5}
& & Dynamic TV & 2.76 & 1.07 & \\ \cline{3-6} 
& & Static HMD & 2.24 & 1.01 & \multirow{2}{*}{0.27 (b)} \\ \cline{3-5}
& & Dynamic HMD & 1.71 & 0.78 & \\ \hline
\multirow{8}{*}{\begin{tabular}[c]{@{}c@{}}Level 4\\ Resolution - 308p\end{tabular}} & \multirow{4}{*}{\begin{tabular}[c]{@{}c@{}}Woman\\ 8.7 KB\end{tabular}} & Static TV & 4.94 & 1.02 & \multirow{2}{*}{0.45 (c)} \\ \cline{3-5}
& & Dynamic TV & 3.71 & 1.30 & \\ \cline{3-6} 
& & Static HMD & 2.53 & 1.47 & \multirow{2}{*}{0.21 (a)} \\ \cline{3-5}
& & Dynamic HMD & 2.00 & 1.00 & \\ \cline{2-6} 
& \multirow{4}{*}{\begin{tabular}[c]{@{}c@{}}City\\ 6.74 KB\end{tabular}}  & Static TV & 3.59 & 1.07 & \multirow{2}{*}{0.29 (b)} \\ \cline{3-5}
& & Dynamic TV & 2.82 & 1.29 & \\ \cline{3-6} 
& & Static HMD & 2.59 & 1.00 & \multirow{2}{*}{0.05 (a)} \\ \cline{3-5}
& & Dynamic HMD & 2.18 & 0.94 & \\ \hline
\multirow{8}{*}{\begin{tabular}[c]{@{}c@{}}Level 5\\ Resolution - 154p\end{tabular}} & \multirow{4}{*}{\begin{tabular}[c]{@{}c@{}}Woman\\ 5.19 KB\end{tabular}} & Static TV & 5.82 & 0.35 & \multirow{2}{*}{0.24 (a)} \\ \cline{3-5}
& & Dynamic TV & 5.41 & 0.8 & \\ \cline{3-6} 
& & Static HMD & 3.76 & 1.19 & \multirow{2}{*}{-0.02 (a)} \\ \cline{3-5}
& & Dynamic HMD & 3.76 & 1.3 & \\ \cline{2-6} 
& \multirow{4}{*}{\begin{tabular}[c]{@{}c@{}}City\\ 4.24 KB\end{tabular}}  & Static TV & 5.35 & 0.74 & \multirow{2}{*}{0.31(b)} \\ \cline{3-5}
& & Dynamic TV & 4.47 & 1.31 & \\ \cline{3-6} 
& & Static HMD & 3.59 & 1.44 & \multirow{2}{*}{0.06 (a)} \\ \cline{3-5}
& & Dynamic HMD & 3.06 & 1.56 & \\ \hline
\multicolumn{6}{|l|}{a: small effect size ($r$=0.10), b: medium effect size ($r$=0.30), c: large effect size ($r$=0.50)} \\ \hline

\end{tabular}}
\label{tab2}
\end{table}

\section{Discussion}

From this experiment we can draw the following three conclusions which are applicable to both the TV and VR scenarios. Firstly, users performed better at the static visual acuity test than at the dynamic test which is in line with past studies \cite{Horng}, \cite{Miller2}. Secondly, dynamic acuity correlated with perception of impairment. People who scored higher in dynamic sight tests, were able to see the differences between the reference image and impaired images during the dynamic impairment rating. Finally, although the results were statistically insignificant in the impairment ratings, we found promising effect sizes (Pearson's correlation) between static and dynamic conditions for the different impairment levels. Unlike significance tests, effect size is independent of sample size. Statistical significance, on the other hand, depends upon both sample size and effect size. For this reason, $p$ values are considered to be confounded because of their dependence on sample size. Sometimes a statistically significant result means only that a huge sample size was used \cite{Ellis,Sullivan}. From all the above, we can assume that the impairment perception of the human visual system in a static and a dynamic environment is linearly correlated, something that can be exploited to create practical QoE bandwidth management solutions.

Future work of this research consists of extending the tests to regular video, $360^o$ video, and light field video. Additionally, a more realistic motion scheme for the user is a work in progress, where the dynamic visual acuity and content evaluation can be measured during actual movement of the subject. That movement can be directly dependent on the viewed content which will lead the subject throughout the process. In other words, the subject will evaluate various VR content while freely roaming in space, which is a more realistic scenario than forced horizontal head movement. Finally, while a link between poor visual system performance and cybersickness was confirmed in the past \cite{Davis}, future extensions of this work could probe whether or not there is a connection between user's visual acuity and cybersickness.

\section{Conclusions}

We investigated the correlation of (dynamic) visual acuity and impairment rating of 2D images on TV and on HMD. Results showed that visual acuity and impairment ratings are indeed correlated, while users' scores on TV and on HMD for the same content are not necessarily correlated. Although using distorted images and maintaining a good user experience might seem mutually exclusive, our experiment shows that, when in motion, users rate certain impairment levels of images as less perceptible than static users. To the best of our knowledge, this study was the first to combine and compare static and dynamic visual acuity and impaired image quality tests both on TV and on HMD. The identified range of quality tipping points can be used as a baseline for subsequent research using a dataset consisting of various types of video content used on HMDs.


\begin{thebibliography}{00}

\bibitem{Langley} H. Langley, \textit{"This is what virtual reality will (probably) look like in 2021"}, Wareable, 2016, accessed at: https://www.wareable.com.
\bibitem{Abrash} M. Abrash, \textit{"Oculus Connect 3 Opening Keynote"}, Oculus Connect 3, 2016, accessed at: https://www.youtube.com/watch?v=AtyE5qOB4gw and http://www.oculusconnect.com
\bibitem{Wallendael} G. Van Wallendael, P. Coppens, T. Paridaens, N. Van Kets, W. Van den Broeck, P. Lambert, (2016). \textit{"Perceptual quality of 4K-resolution video content compared to HD"}. 2016 Eighth International Conference on Quality of Multimedia Experience (QoMEX), Lisbon, 2016, pp. 1-6.
\bibitem{Davis} S. Davis, K. Nesbitt and E. Nalivaiko, \textit{"A systematic review of Cybersickness"}, in Proceedings of the Conference on Interactive Entertainment, pp. 1-9, 2014.
\bibitem{Sue} S. V. G. Cobb, S. Nichols, A. Ramsey and J. R. Wilson, \textit{"Virtual Reality-Induced Symptoms and Effects (VRISE)"}, Presence: Teleoperators and Virtual Environments, Volume~8, Issue~2, pp.~169–186, 1999.
\bibitem{Kennedy} R. S. Kennedy , N. E. Lane , K. S. Berbaum  and M. G. Lilienthal, \textit{"Simulator Sickness Questionnaire: An enhanced method for quantifying simulator sickness"}, International Journal of Aviation Psychology, Volume~3, Issue~3, pp.~203–220, 1993.
\bibitem{Patney} A. Patney, M. Salvi, J. Kim, A. Kaplanyan, C. Wyman, N. Benty, D. Luebke and A. Lefohn, \textit{"Towards foveated rendering for gaze-tracked virtual reality"}, in Siggraph Asia, Macao, 2016.
\bibitem{Horng} C. T. Horng, Y. S. Hsieh, M. L. Tsai, W. K. Chang, T. H. Yang, C. H. Yauan, C. H. Wang, W. H. Kuo and Y. C. Wu, \textit{"Effects of horizontal acceleration on human visual acuity and stereopsis"}, in International Journal of Environmental Research and Public Health, vol.~12, no.~1, pp.~910-926, 2015. 
\bibitem{Demer} J. L. Demer and F. Amjadi, \textit{"Dynamic Visual-Acuity of Normal Subjects During Vertical Optotype and Head Motion"}, Investigative opthalmology \& visual science, vol.~34, no.~6, pp.~1894-1906, 1993.
\bibitem{Miller} J. W. Miller, and E. Ludvigh, \textit{"Study of Visual Acuity during Ocular Pursuit of Moving Test Objects. I. Introduction"}, in Journal of the Optical Society of America, vol.~48, no.~11, pp.~799-802, 1958.
\bibitem{Sloan} \textit{"Sloan Letter Revised Series ETDRS Charts (3 Meter)"}, Precision Vision, Accessed on December 6, 2019 from http://www.precision-vision.com/product/sloan-letter-revised-series-etdrs-charts-3-meter/.

\bibitem{Peters} B. T. Peters, A. P. Mulavara, H. S. Cohen, H. Sangi-Haghpeykar and J. J. Bloomberg, \textit{"Dynamic visual acuity testing for screening patients with vestibular impairments"}, in the Journal of Vestibular Research, Volume~22, Issue~2, pp.~145-151, 2012.
\bibitem{Li} C. Li, J. L. Beaumont, R. M. Rine, J. Slotkin and M. C. Schubert, \textit{"Normative Scores for the NIH Toolbox Dynamic Visual acuity Test from 3 to 85 Years"}, in Frontiers in Neurology, Volume~5, pp.~223, 2014.

\bibitem{Anstis} S. M. Anstis, \textit{"A chart demonstrating variations in acuity with retinal position"}, in Vision Research, Volume~14, Issue~7, pp.~589-592, ISSN~0042-698, 1974.
\bibitem{Monti} P. W. Monti, \textit{"Lateral Masking of end Elements by Inner Elements in Tachistoscopic Pattern Perception"}, in Perceptual and Motor Skills, Volume~36, Issue~3, pp.~777-778, 1973.
\bibitem{Fischer} B. Fischer and E. Ramsperger, \textit{"Human express saccades: extremely short reaction times of goal directed eye movements"}, in Experimental Brain Research, Volume~57, Issue~1, pp.~191-195, 1984.
\bibitem{Schutz} A. C. Schutz, D. I. Braun and K. R. Gegenfurtner, \textit{"Object recognition during foveating eye movements"}, in Vision Research, Volume~49, Issue~18, pp.~2241-2253, ISSN~0042-6989, 2009.
\bibitem{Brown} B. Brown, \textit{"Dynamic visual acuity, eye movements and peripheral acuity for moving targets"}, in Vision Research, Volume~12, Issue~2, pp.~305-321, ISSN~0042-6989, 1972.
\bibitem{Demer2} J. L. Demer, V. Honrubia and  R. Baloh, \textit{"Dynamic visual acuity: a test for oscillopsia and vestibulo-ocular reflex function"},  in the American Journal of Otology, Volume~15, Issue~3, pp.~340-347, 1994.
\bibitem{Doumanoglou} A. Doumanoglou et al., \textit{"Quality of Experience for 3-D Immersive Media Streaming"}, in IEEE Transactions on Broadcasting, Volume~64, no.~2, pp.~379-391, 2018.
\bibitem{Recio} R. Recio, P. Carballeira, J. Gutierrez and N. Garcia, \textit{"Subjective Assessment of Super Multiview Video with Coding Artifacts"}, in IEEE Signal Processing Letters, Volume~24, no.~6, pp.~868-871, 2017.
\bibitem{Ling} S. Ling, J. Gutierrez, G. Ke and P. Le Callet, \textit{"Prediction of the Influence of Navigation Scan-path on Perceived Quality of Free-Viewpoint Videos"}, in  Computer Vision and Pattern Recognition, abs/1810.04409, 2018.
\bibitem{Gutierrez} J. Gutierrez, E. J. David, A. Coutrot, M. P. Da Silva and P. L. Callet, \textit{"Introducing UN Salient360! Benchmark: A platform for evaluating visual attention models for 360° contents"}, in the Tenth International Conference on Quality of Multimedia Experience (QoMEX), pp.~1-3, 2018.
\bibitem{Sitzmann} V. Sitzmann et al., \textit{"Saliency in VR: How Do People Explore Virtual Environments?"}, in IEEE Transactions on Visualization and Computer Graphics, Volume~24, no.~4, pp.~1633-1642, 2018.

\bibitem{Rine} R. M. Rine and J. Braswell, \textit{"A clinical test of dynamic visual acuity for children"}, in International Journal of Pediatric Otorhinolaryngology, vol.~67, no.~11, pp.~1195-1201, 2003.
\bibitem{Keimel} C. Keimel, \textit{"Design of Video Quality Metrics with Multi-Way Data Analysis"}, Springer Singapore, 2016.
\bibitem{ITU-T} ITU-T, \textit{"Subjective video quality assessment methods for multimedia applications"}, 2009.

\bibitem{ITU} ITU, \textit{Studies toward the unification of picture assessment methodology}, 1990.
\bibitem{ITU-R} ITU-R BT.500-13, \textit{"Methodology for the subjective assessment of the quality of television pictures"}, 2012.
\bibitem{MOS} ITU-R P.800.1, \textit{"Mean opinion score (MOS) terminology"}, 2016.
\bibitem{Miller2} J. W. Miller, \textit{"Study of Visual Acuity during the Ocular Pursuit of Moving Test Objects. II. Effects of Direction of Movement, Relative Movement, and Illumination"}, Journal of the Optical Society of America, vol.~48, no.~11, pp.~803-808, 1958.
\bibitem{Ellis} P. D. Ellis, \textit{"The Essential Guide to Effect Sizes: Statistical Power, Meta-Analysis, and the Interpretation of Research Results"}, Cambridge University Press, doi:10.1017/CBO9780511761676, 2010.
\bibitem{Sullivan} G. M. Sullivan and R. Feinn, \textit{"Using Effect Size—or Why the P Value Is Not Enough
"}, in Journal of Graduate Medical Education, vol.~4, no.~3, pp.~279-282, 2010.

\end{thebibliography}
\end{document}